\documentclass[aps,prx,twocolumn,superscriptaddress,floatfix,nofootinbib]{revtex4-2}
\usepackage{amsmath,amssymb,bm}
\usepackage{graphicx}
\usepackage{booktabs}
\usepackage[colorlinks=true,linkcolor=blue,citecolor=blue,urlcolor=blue]{hyperref}
\usepackage{xcolor}

\newcommand{\dd}{\mathrm{d}}
\newcommand{\kB}{k_\mathrm{B}}
\newcommand{\vol}{\mathrm{vol}}
\newcommand{\el}{\mathrm{el}}
\newcommand{\surf}{\mathrm{surf}}
\newcommand{\eff}{\mathrm{eff}}
\newcommand{\tens}{\mathrm{tens}}
\newcommand{\bend}{\mathrm{bend}}

\newcommand{\hmo}{h_{-1}}
\graphicspath{{./}}

\usepackage{xr-hyper}          
\usepackage{hyperref}
\begin{document}

\title{$1/f$ frequency-noise reduction in nanomechanical resonators by dissipation dilution}


\author{Kostas Kanellopulos}
\affiliation{Institute of Sensor and Actuator Systems, TU Wien, Gusshausstrasse 27-29, 1040 Vienna, Austria}
\author{Silvan Schmid}
\email{silvan.schmid@tuwien.ac.at}
\affiliation{Institute of Sensor and Actuator Systems, TU Wien, Gusshausstrasse 27-29, 1040 Vienna, Austria}

\date{\today}

\begin{abstract}
The frequency noise of a resonant nanomechanical sensor is well understood down to its thermomechanical limit, but the longest useful averaging time is set by a flicker, or $1/f$, frequency-noise floor whose origin has remained unknown. We show that this floor is the thermal noise of the tension. By the fluctuation-dissipation theorem, the tension of a resonator with fixed ends fluctuates by an amount set by the loss angle of its material; for structural damping, these fluctuations produce $1/f$ frequency noise with a flicker coefficient $\hmo=E\kB T/(2\pi V\sigma_0^2Q_\el)$, with $Q_\el$ the intrinsic loss under elongation, that contains no free parameter and is diluted by the tensile stress $\sigma_0$ in the same way as damping is in dissipation dilution. Silicon nitride strings and membranes from 30 to 340~nm thickness follow the predicted volume scaling, the predicted $\sigma_0^{-2}$ dilution across resonators and within stress-engineered strings, and the predicted thickness dependence, which places noise-active two-level systems in the volume and at the surfaces of the resonators, all at the absolute level set by the independently measured intrinsic loss. The same relation accounts for published data between 78 and 296~K; below 78~K, the loss measured at the resonance no longer represents the loss probed by the noise, marking the limit of the structural-damping assumption. Dissipation dilution therefore lowers the $1/f$ floor, and the longest useful integration time of a resonant sensor is set by its prestress, volume, temperature, and intrinsic loss angle.

\end{abstract}

\maketitle

\subsection*{Introduction}
Nanomechanical resonators are among the most sensitive transducers available \cite{bachtold2022mesoscopic}, detecting mass from single proteins and nanoparticles to single atoms \cite{hanay2012single,neumann2024nanomechanical,yang2006zeptogram,chaste2012nanomechanical,gil2010nanomechanical,schmid2013real}, force gradients from spins and charges \cite{garner2004force,BOUCHER2026108165,visani2026near,cleland1998nanometre,Unterreithmeier2009Universal}, temperature changes in the microkelvin range \cite{larsen2013micro,larsen2011ultrasensitive,ferreiro2021micro}, and absorbed power from single molecules or electromagnetic radiation down to the femtowatt level \cite{larsen2013photothermal,schmid2014low,piller2022thermal,martini2025uncooled}. In every case, the signal is a shift of the resonance frequency, and the resolution is set by the frequency stability of the resonator, quantified by the Allan deviation $\sigma_y(\tau)$ at the integration time $\tau$ of the measurement \cite{Cleland2002,Schmid2023}. Frequency stability is equally central to self-sustained nano- and micromechanical oscillators, which serve as compact frequency references \cite{feng2008self,kim2007frequency,van2012review}.

Ideally, $\sigma_y$ is limited by the thermomechanical noise of the resonator and the detection noise of the transducer, both of which enter as phase noise and are converted into frequency noise by the phase slope $2Q/\omega_0$ \cite{Besic2023,Demir2021,Schmid2023}. Thermomechanical noise gives $\sigma_y\propto\tau^{-1/2}$; detection noise, amplified in the closed loops needed to track high-$Q$ resonators, gives $\sigma_y\propto\tau^{-1}$ \cite{Demir2021,Schmid2023}. Both average down with $\tau$. In large tensioned strings and membranes, whose frequency responds strongly to temperature, photothermal back-action from the readout \cite{Sadeghi2020} and the statistical temperature fluctuations of the resonator \cite{zhang2023demonstration,Schmid2023} add noise at long $\tau$; both can be reduced by design or by dual-mode noise-cancellation schemes \cite{kharbanda2026chip}, which can reduce correlated frequency noise, including flicker noise.

Flicker frequency noise typically sets the ultimate limit on frequency stability \cite{bachtold2022mesoscopic}. It is the component of the frequency noise with a $1/f$ spectrum, $S_y(f)=\hmo/f$, whose Allan variance is flat, $\sigma_y^2=2\ln2\,\hmo$ \cite{Roukes2026}; we refer to it as flicker noise or $1/f$ noise interchangeably, and to $\hmo$ as the flicker coefficient. The flicker plateau sets the longest useful integration time and hence the lowest Allan deviation a sensor can reach; averaging beyond it yields no gain, nor does a stronger drive, a better transducer, or a quieter laser. Cleland and Roukes anticipated a floor from the reorientation of defects with a broad distribution of relaxation times \cite{Cleland2002}; Fong \emph{et al.} resolved it in ultrahigh-$Q$ silicon nitride strings and found a spectrum scaling as $k_BT/f$ \cite{Fong2012}; Sansa \emph{et al.} showed that no published device reaches its thermomechanical limit, the gap being two orders of magnitude on average, and that the excess is a fluctuation of the resonance frequency itself \cite{Sansa2016}. Most recently, Maksymowych \emph{et al.} resolved the constituents of this ensemble one at a time: in phononic-crystal resonators at millikelvin temperatures, the resonance frequency switches between discrete values as individual two-level defects change state \cite{Maksymowych2025Frequency}.

The most complete account since Sansa \emph{et al.} \cite{Sansa2016}, is a recent survey of 66 devices spanning 13 decades in volume, which established that the product $\hmo V$ is invariant across materials, geometries and transduction schemes, and thereby identified the fluctuators as material defects at a density that is remarkably universal \cite{Roukes2026}. 
Two questions could not be settled from the published record. Because reported devices are geometrically self-similar, the data cannot distinguish fluctuators in the volume from fluctuators at the surfaces.  
And the record does not show whether the flicker noise floor depends on prestress, which a controlled-stress series could determine, but such a series had not been performed. Here, we show from the fluctuation-dissipation theorem (FDT), without a microscopic model, that in a resonator with fixed ends, the floor is diluted by the prestress.


\subsection*{The flicker floor as thermal noise of the tension}

The relation between the flicker floor and the dissipation follows from the fluctuation-dissipation theorem (FDT) \cite{Callen1951}. The ends of a tensioned resonator are fixed, so its elongation is preset, and the quantity free to fluctuate is the tension. For a body of volume $V$ under uniform stress, the FDT gives the one-sided spectral density with units [Pa$^2$/Hz] of the stress as
\begin{equation}
S_\sigma(\omega)=\frac{4\kB T}{\omega}\,\mathrm{Im}\,\hat K(\omega),
\label{eq:fdtsigma}
\end{equation}
with the frequency $\omega=2\pi f$ and the response function $\hat K$, which is a generalized stiffness of the stress to its work-conjugate coordinate, the volume-integrated strain $\partial W/\partial\sigma=V\varepsilon$. With the stored elastic energy $W=\tfrac12E\varepsilon^2V=\tfrac12\sigma^2V/E$, the elastic stiffness of a string, whose transverse strain is free, is $[(\partial W/\partial\sigma)/\sigma]^{-1}=E/V$. A membrane is constrained in both in-plane directions, which changes the coupling of a two-level system (TLS) to the tension and to the bending strain by the same factor; the two nearly cancel, leaving a prefactor of $1/(1-\nu^2)=1.06$ for $\nu=0.23$ (Section~\ref{SI-sec:relations}), which we set to unity. By the correspondence principle of linear viscoelasticity, the response in the frequency domain follows from the elastic one by replacing $E$ with the complex modulus,
\begin{equation}
\hat K(\omega)=\frac{E^*(\omega)}{V},
\label{eq:chi}
\end{equation}
where $E^*=E+iE''=E[1+i\phi(\omega)]$ is the complex Young's modulus and $\phi=E''/E$ the loss angle of the material (the loss tangent $\tan\delta$ of materials science), the same quantity that sets the intrinsic loss in dissipation dilution. Hence $\mathrm{Im}\,\hat K= E\phi/V$ and $S_\sigma=4\kB TE\phi/(\omega V)$: the stress of a larger body fluctuates less. The resonance frequency is determined by tension, and the fractional frequency change is $y=\delta\sigma/2\sigma_0$, so that  $S_y=S_\sigma/(4\sigma_0^2)$, or
\begin{equation}
S_y(f)=\frac{\kB T\,E\,\phi(f)}{\omega\,\sigma_0^2V}=\frac{E\kB T\,\phi(f)}{2\pi\sigma_0^2V}\,\frac{1}{f}.
\label{eq:fdt}
\end{equation}
The spectral shape of the fractional frequency noise is set by the frequency dependence of the loss angle. Viscous damping, $E''\propto\omega$, gives white frequency noise; a single relaxation process, such as thermoelastic damping, gives a Lorentzian that is white below its relaxation rate \cite{zhang2023demonstration, Schmid2023}; only structural damping, a loss angle independent of frequency, gives $S_y\propto1/f$. The flicker floor is therefore the signature of structural damping at frequencies $f\sim1/(2\tau)$ sampled by the Allan deviation with integration times $\tau$ in the range from mHz to a few Hz, with $\hmo V\sigma_0^2=E\kB T\phi/(2\pi)$. 
Structural damping is the characteristic loss of amorphous solids, such as silicon nitride \cite{fedorov2018evidence}. It arises from a broad, log-uniform distribution of relaxation rates of TLS, atomic configurations that switch between two states by thermal activation over a barrier or by tunneling through it \cite{Phillips1987,Dutta1981}. TLS have been proposed as the dominant source of friction in nanomechanical resonators \cite{seoanez2008surface,gruenke2025surface}. 

The flicker floor is thus the thermal noise of the tension, in the same sense in which the thermomechanical limit is the thermal noise of the mode displacement, and the parallel to dissipation dilution is exact: the same loss angle sets $Q^{-1}=\phi\,W_\bend/W_\tens$ \cite{Schmid2023} and $\hmo=\kB TE\phi/(2\pi\sigma_0^2V) = \phi\,[\kB T/(4\pi)]/W$, divided in both by the tensile energy stored in the resonator.

If the loss angle is structural over the eight decades separating the low-frequency Allan band from the resonance, it can be taken from the quality factor measured at $\omega_0$. Two qualifications fix which quality factor. First, only the loss of the material enters: external channels that damp the mode, such as radiation into the substrate or gas damping, add thermomechanical noise but no tension noise, so the relevant quantity is the intrinsic $Q_\mathrm{int}$ obtained from dissipation dilution, not the total $Q$. Second, $Q_\mathrm{int}$ is measured under the bending strain of the mode, whereas Eq.~\eqref{eq:fdt} needs the loss under uniform elongation, $\phi=1/Q_\el$, so that
\begin{equation}
\hmo=\frac{E\kB T}{2\pi V\sigma_0^2}\,\frac{1}{Q_\el}.
\label{eq:main}
\end{equation}
Eq.~\eqref{eq:main} contains no free parameter (all material parameters in Table~\ref{tab:params}, Methods). The two losses differ where the loss is confined to the surfaces: there, the bending surface loss $1/Q_\surf=1/(\beta h)=6 h_\mathrm{S}/(Q_\mathrm{S} h)$ for a dissipative surface film of thickness $h_\mathrm{S}$ and loss $1/Q_\mathrm{S}$ \cite{Yasumura2000,Villanueva2014} enters the elongation loss as $2h_\mathrm{S}/(Q_\mathrm{S} h) = 1/(3Q_\surf)$, provided the surface loss is itself structural. With the volume loss, which enters both equally,
\begin{equation}
\frac{1}{Q_\el}=\frac{1}{Q_\vol}+\frac{1}{3\beta h},\quad\text{against}\quad \frac{1}{Q_\mathrm{int}}=\frac{1}{Q_\vol}+\frac{1}{\beta h},
\label{eq:Qel}
\end{equation}
so that the floor is flat above and rises as $1/h$ below the crossover thickness $h^*=Q_\vol/(3\beta)$.

The structural-damping hypothesis spans eight decades in frequency. The noise at $f$ and the loss at $\omega_0$ are each carried by the TLS whose correlation time matches the measurement, $\tau\approx1/\omega$, and each responds to the TLS within about one decade of correlation time around that value, the width of the Debye window $\omega\tau/(1+\omega^2\tau^2)$ at fixed $\omega$ \cite{Dutta1981}. For thermally activated relaxation, $\tau=\tau_0\exp(E_b/\kB T)$ \cite{buchenau2001mechanical}, with a microscopic time $\tau_0$ of the order of  ($\sim10^{-12}$~s) \cite{Fong2012,buchenau2001mechanical}, the two windows are separated in activation energy by
\begin{equation}
\Delta E_b=\kB T\ln\frac{\omega_0}{2\pi f},
\label{eq:window}
\end{equation}
independent of $\tau_0$; for $\omega_0/2\pi\sim1$~MHz and $f\sim0.1$~Hz this is $16\,\kB T$, about $0.4$~eV at room temperature, which is the separation on which Ref.~\cite{Roukes2026} based the expectation that the two bands are decoupled. Eq.~\eqref{eq:fdt} asks for the loss angle at $f$. In glasses, loss measured from hertz to gigahertz has been shown to sample a single, temperature-independent barrier distribution \cite{buchenau2001mechanical}. Replacing it by the loss angle measured at $\omega_0$ requires that the density of TLS per decade of $\tau$, which $S_y\propto1/f$ requires to be flat within the Allan band and a frequency-independent $Q$ requires to be flat around $\omega_0$, is also flat across the interval of Eq.~\eqref{eq:window} between them. The relation links the ensemble densities in the two windows, not individual TLS, and the separation of the windows enters only through this condition. Whether it is met is an empirical question, which the temperature dependence addresses below. 

Eq.~\eqref{eq:fdt} makes the stress dependence general. Whatever the microscopic coupling, the tension noise of a body in thermal equilibrium is fixed by its loss angle, so that $\hmo\propto\sigma_0^{-2}$ at fixed $\phi$. A fluctuator can reach the resonance frequency in two ways. With fixed ends, it shifts the frequency through the tension, $y=\delta\sigma/2\sigma_0$. First, a change in local strain $\delta\varepsilon$ gives $\delta\sigma=E\,\delta\varepsilon$ and $y=E\,\delta\varepsilon/2\sigma_0$, diluted by the prestress; second, a change in local modulus $\delta E$ at the prestrain $\varepsilon_0=\sigma_0/E$ gives $\delta\sigma=\delta E\,\varepsilon_0$ and $y=\delta E\,\varepsilon_0/2\sigma_0=\delta E/2E$, independent of it. The latter is consistent with Eq.~\eqref{eq:fdt} because the loss of the modulus coupling grows as $\sigma_0^2$ under prestress and cancels the dilution (Section~\ref{SI-sec:micro-model}). The two couplings, therefore, differ in the stress dependence of the intrinsic loss. In silicon nitride, the intrinsic loss does not depend on composition \cite{Villanueva2014} and hence not on the prestress that composition sets, which identifies strain coupling as the dominant one, a conclusion we test directly at fixed composition below.

\begin{figure}
\centering
\includegraphics[width=\columnwidth]{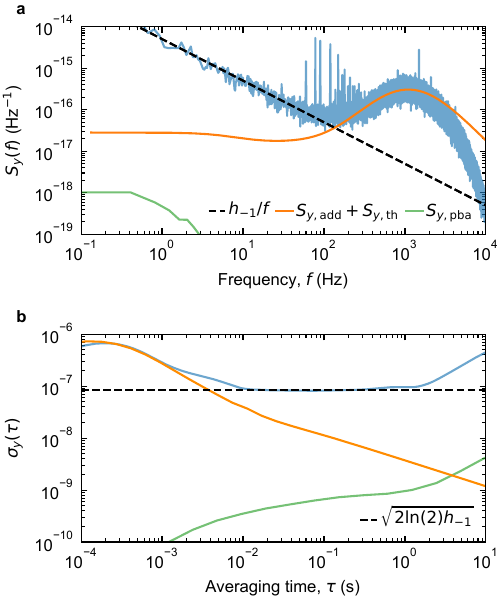}
\caption{Flicker noise measurements. \textbf{a}, Fractional frequency-noise power spectral density $S_y(f)$ of a $56\ \mathrm{nm}$ thick, $5\ \mathrm{\mu m}$ wide, $100\ \mathrm{\mu m}$ long string with a prestress of $380\ \mathrm{MPa}$. Blue: measured spectrum; orange: combined additive phase noise and temperature fluctuations; green: frequency photothermal back-action; dashed black: $1/f$ flicker frequency noise, $S_y(f)=h_{-1}f^{-1}$. \textbf{b}, Corresponding Allan deviation. The flicker-noise contribution produces the plateau indicated by the dashed line, $\sigma_y=\sqrt{2\ln2\hmo}$.}
\label{fig:setup}
\end{figure}

Beyond the inverse volume dependence, the relation (\ref{eq:main}) predicts a $\sigma_0^{-2}$ dilution of the floor, a thickness dependence that differs between the two placements, and a linear temperature dependence. We test the model predictions on silicon nitride (SiN) strings and membranes spanning 30 to 340~nm in thickness, with a prestress in the range of $100$~MPa to $1$~GPa (Table~\ref{tab:samples}, Methods), whose intrinsic loss channels are quantitatively established (see Section~\ref{SI-sec:Qfit}) \cite{Villanueva2014,Benga2026,Cavalleri2026} , together with published $1/f$ data on SiN resonators from \cite{Fong2012}. Fig.~\ref{fig:setup}(a) shows the fractional frequency-noise spectrum of a representative string: below about 100~Hz, the additive, temperature-fluctuation, and back-action contributions fall away, and the spectrum settles on the $S_y(f)=\hmo/f$ law. In the Allan deviation, Fig.~\ref{fig:setup}(b), the same term appears as the flat floor $\sigma_y=\sqrt{2\ln2\,\hmo}$ that no further averaging lowers (see Methods).  

\subsection*{Volume scaling}

\begin{figure}
\centering
\includegraphics[width=\columnwidth]{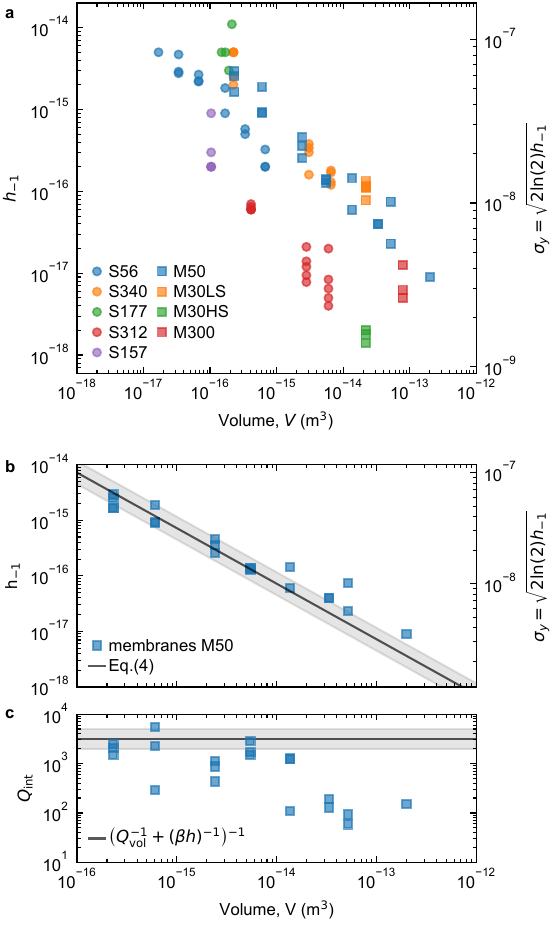}
\caption{Volume scaling of the flicker coefficient. \textbf{a}, Flicker coefficient $\hmo$ as a function of resonator volume $V$ for silicon nitride strings and membranes measured in this work (see Table~\ref{tab:samples} in Methods). \textbf{b}, $\hmo$ as a function of $V$ of membranes from a single batch. \textbf{c}, Corresponding intrinsic quality factor $Q_\mathrm{int}$ as a function of $V$. The solid lines show the parameter-free predictions with the parameters of Table~\ref{tab:params} (no fit). The shaded region denotes the $95\%$ scatter band (see Section~\ref{SI-sec:Qfit}).}
\label{fig:volume}
\end{figure}

Fig.~\ref{fig:volume}(a) shows $\hmo$ against the resonator volume $V$ for all devices, confirming the inverse correlation demonstrated by \cite{Sansa2016,Roukes2026} as predicted by (\ref{eq:fdt}). Across the data set, the points scatter over two decades at a given volume, ordered by stress. Fig.~\ref{fig:volume}(b) singles out a set of 50~nm thick membranes. Fabricated in a single batch, they share the same nominal prestress, thickness, and stoichiometry, so their lateral size is the only variable parameter. Their flicker coefficient accurately follows the parameter-free prediction for membranes (\ref{eq:main}), using the intrinsic loss as a measure for the fluctuator density.

The membrane series also separates the measured quality factor from the loss angle that enters Eq.~\eqref{eq:fdt}. The intrinsic quality factors obtained from dissipation dilution, $Q_\mathrm{int}=Q/D$ (see Methods), span up to one order of magnitude between membranes of the same size; the smaller membranes reach the intrinsic ceiling of a 50~nm film (Table~\ref{tab:params}), the others lie below it by clamping loss, the coupling of the fundamental mode to the modes of chip and mount \cite{wilson2011high,Villanueva2014,Benga2026}. This loss does not dilute and is not a loss of the material. Had the noise floor followed $Q_\mathrm{int}$, membranes of equal size would differ in $\hmo$ by the same factor as in $Q_\mathrm{int}$; instead, their flicker noise floors agree within a factor of $\sim 2$ [Fig.~\ref{fig:volume}(b)]. The flicker noise floor is tied to the material's loss angle, not to the measured $Q$, which includes channels that damp the mode without dissipating the tension.

\subsection*{Dilution with tensile stress}
\begin{figure}
\centering
\includegraphics[width=\columnwidth]{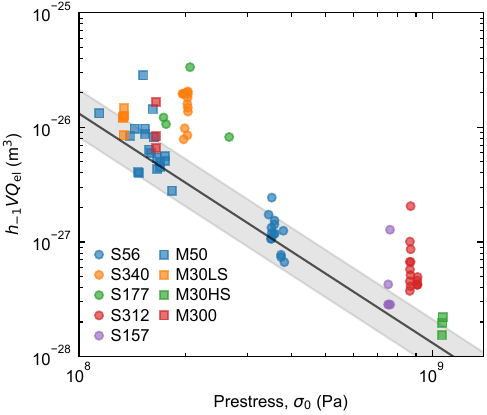}
\caption{Stress dilution of the flicker floor. Flicker invariant, $h_{-1}\cdot V \cdot Q_{el}$ as a function of prestress $\sigma_0$ for all samples measured in this work (see Table~\ref{tab:samples} in Methods). The solid line shows the parameter-free prediction with the parameters of Table~\ref{tab:params} (no fit), while the shaded region denotes the $95\%$ scatter band (see Section~\ref{SI-sec:Qfit}).}
\label{fig:stress}
\end{figure}

Normalized with the resonator volume and material loss,
Fig.~\ref{fig:stress} shows $\hmo V$ against $\sigma_0$ for all strings and membranes. The data follow $\hmo V Q_\el\propto\sigma_0^{-2}$, as predicted by Eq.~(\ref{eq:main}), over one decade in $\sigma_0$, with our silicon-rich and stoichiometric devices on one line. The model sets a lower bound on the floor. The strings of samples S340, S177, S312, and S157 consistently lie above it on average. These samples share a frame design with a centimeter-long through-hole that reduces the stiffness of the silicon  frame~\cite{schmid2011damping}. A compliant frame could add frequency noise in two ways: by transmitting mount and thermal strain $\varepsilon_\mathrm{f}$ to the string tension, $\delta y = E\,\delta\varepsilon_\mathrm{f}/(2\sigma_0)$, and by lowering the frequencies of the frame modes to which the string couples. Neither contribution would scale with the string volume, and the first would follow the same $\sigma_0^{-2}$ dependence as the intrinsic floor, consistent with the excess running parallel to the model line. The general trend nevertheless shows that stress-induced dissipation dilution lowers the flicker floor, the counterpart of the
enhancement of the quality factor by the same mechanism~\cite{schmid2011damping,Schmid2023,verbridge2006high,zwickl2008high,engelsen2024ultrahigh}.

Because the prestress of silicon nitride is set by its composition, the $\sigma_0^{-2}$ trend of Fig.~\ref{fig:stress} could in principle reflect a composition dependence of the fluctuator density. Two observations argue against this. First, the intrinsic loss, which in the model is set by the same fluctuator density, does not depend on composition \cite{Villanueva2014}. Within our data set, this is illustrated by the 30~nm membranes M30HS and M30LS. They have the same geometry and volume but different composition, stoichiometric and silicon-rich, and hence different prestress, 1070 and 134~MPa. Their intrinsic loss falls in the same range, $Q_\mathrm{int}^{-1}=(0.3\text{--}1.5)\times10^{-3}$ [Fig.~\ref{fig:thickness}(b)], whereas their flicker coefficients differ by a factor of about 60 [Fig.~\ref{fig:volume}(a)], as predicted by $(\sigma_{0,\mathrm{HS}}/\sigma_{0,\mathrm{LS}})^2\approx64$. Second, $\hmo V$ plotted against the Si/N ratio follows the variation in stress, whereas $\hmo V\sigma_0^2$ does not depend on it (Section~\ref{SI-sec:stoich}). Composition, therefore, enters through the prestress, not through the fluctuator density.

\subsection*{Stress tuning}

\begin{figure}
\centering
\includegraphics[width=\columnwidth]{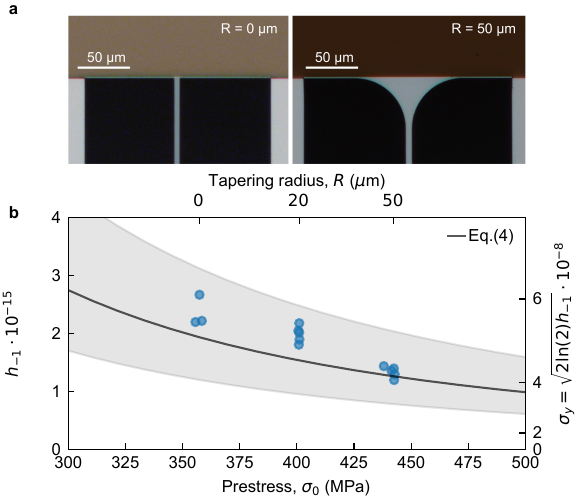}
\caption{Flicker noise reduction through stress tuning. \textbf{a}, Optical micrographs of the stress-engineered strings, featuring clamps with tapering radii $R=0$ and $50\, \mathrm{\mu m}$. \textbf{b}, flicker coefficient $\hmo$ as a function of the effective prestress $\sigma_0$ for 56~nm strings of length $L=200\, \mathrm{\mu m}$ from the same chip, with clamp tapering radii $R=0$, 20 and 50~$\mu$m; the prestress $\sigma_0$ of each device is inferred from its resonance frequency. The solid line shows the parameter-free prediction from Eq.~\eqref{eq:main} (no fit) evaluated with the parameters of Table~\ref{tab:params}, using the nominal string volume and excluding the tapered-clamp regions. The shaded region denotes the $95\%$ scatter band (see Section~\ref{SI-sec:Qfit}).}
\label{fig:tuning}
\end{figure}

The stress series of Fig.~\ref{fig:stress} compares films of different composition, in which the prestress and the fluctuator density could, in principle, as discussed, co-vary. To vary the stress alone, we analyzed strings with tapered clamping points fabricated on one chip from the same 56~nm film (see Fig.~\ref{fig:tuning}a). A tapered clamp of radius $R$ concentrates the film stress in the narrow section of the string \cite{sadeghi2019influence}, so that the fluctuator population, the thickness, and the composition are identical while the tension is not. Extracted from the eigenfrequencies, the prestress rises from 357~MPa without tapering ($R=0$) to 400~MPa for $R=20~\mu$m and 440~MPa for $R=50~\mu$m (Fig.~\ref{fig:tuning}b). Over this range, the flicker coefficient falls from $(2.2$ to $2.7)\times10^{-15}$ to $(1.2$ to $1.45)\times10^{-15}$. Between the medians, this is a factor of 1.6 for a 23\% increase in prestress, against the factor $(440/357)^2=1.52$ predicted by Eq.~\eqref{eq:main}. The three groups are separated well beyond their device-to-device scatter. The dilution of the flicker noise floor with tensile stress is thus observed within a single film, where it cannot be attributed to composition. The model line (\ref{eq:main}) reproduces both the slope and the absolute level.

\subsection*{Thickness dependence}

\begin{figure}
\centering
\includegraphics[width=\columnwidth]{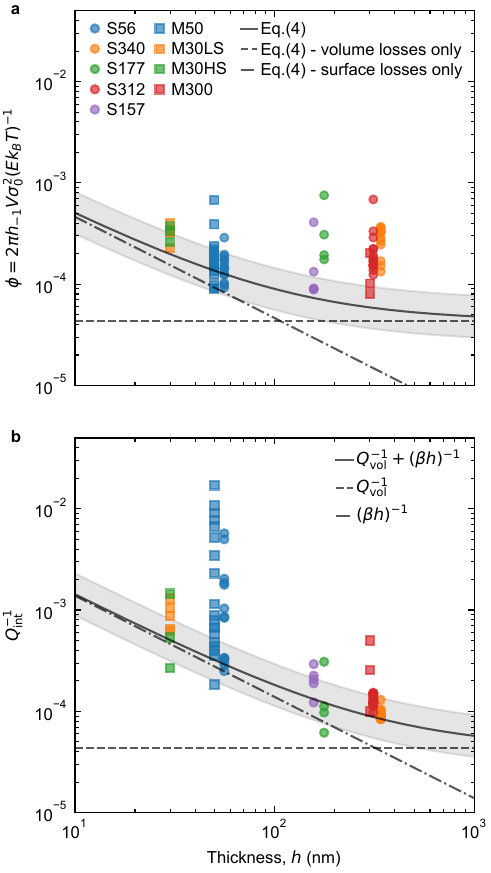}
\caption{Thickness dependence of the flicker floor and of the intrinsic loss. \textbf{a}, Loss angle inferred from the flicker noise, $\phi=2\pi\hmo V\sigma_0^2/(E\kB T)$, against film thickness $h$ for all samples (see Table~\ref{tab:samples} in Methods). Solid line: $1/Q_\el=1/Q_\vol+1/(3\beta h)$; dashed: volume term only; dash-dotted: surface term only. \textbf{b}, Measured intrinsic loss angle $1/Q_\mathrm{int}=D/Q$ of the same devices (Methods). Solid line: $1/Q_\mathrm{int}=1/Q_\vol+1/(\beta h)$; dashed and dash-dotted: the two terms. Both panels use the parameters of Table~\ref{tab:params} without adjustment; shaded bands: 95\% scatter band of the fit in Supplementary Section~\ref{SI-sec:Qfit}.}
\label{fig:thickness}
\end{figure}

With the stress and volume dependence removed, Eq.~\eqref{eq:main} can be inverted to give the loss angle that the flicker noise implies, $\phi=2\pi\hmo V\sigma_0^2/(E\kB T)$, which the model identifies with the elongation loss $\phi=1/Q_\el$. Fig.~\ref{fig:thickness}(a) shows this quantity against thickness for all devices, together with the parameter-free prediction $1/Q_\el=1/Q_\vol+1/(3\beta h)$ and its two limits: flat at $1/Q_\vol$ for thick films and rising as $1/h$ below $h^*=Q_\vol/(3\beta)=107\pm21$~nm. Fig.~\ref{fig:thickness}(b) shows the measured loss angle $1/Q_\mathrm{int}$ of the same devices with the corresponding prediction $1/Q_\vol+1/(\beta h)$ \cite{Villanueva2014} (see Section~\ref{SI-sec:Qfit}), in which the surface term dominates, and the crossover lies at $Q_\vol/\beta\approx320$~nm. Only the lower envelope of the measured loss is intrinsic, since clamping loss adds to the loss of the large membranes and the thin strings without entering the model.

The loss angle $\phi(h)$ inferred from the measured noise lies above the volume-only level of Eq.~(\ref{eq:main}), with an observable increase for the thinnest 30~nm films. The thickness series is thus consistent with noise-active TLS residing both in the volume and at the surfaces of the film, with the floor of thick films set by the volume loss and that of thin films ($h<h^*\approx100$~nm) increasingly by the surface loss that dominates $Q_\mathrm{int}$ below about 300~nm. The weaker rise of the noise, with its crossover at $h^*$ rather than at $Q_\mathrm{vol}/\beta$, reflects the factor of three by which bending overweights the surfaces relative to elongation. The thicker strings (S340, S177, S312, and S157) consistently lie above the model. As discussed with Fig.~\ref{fig:stress}, these samples share a compliant frame design, which could add a volume-independent contribution to their floor. If so, the flat appearance of $\phi(h)$ above 150~nm would reflect this extrinsic contribution rather than the intrinsic loss angle, and the comparison with the model would rest mainly on the thinner strings and the membranes, which have different frame designs.


\subsection*{Temperature dependence}

\begin{figure}
\centering
\includegraphics[width=\columnwidth]{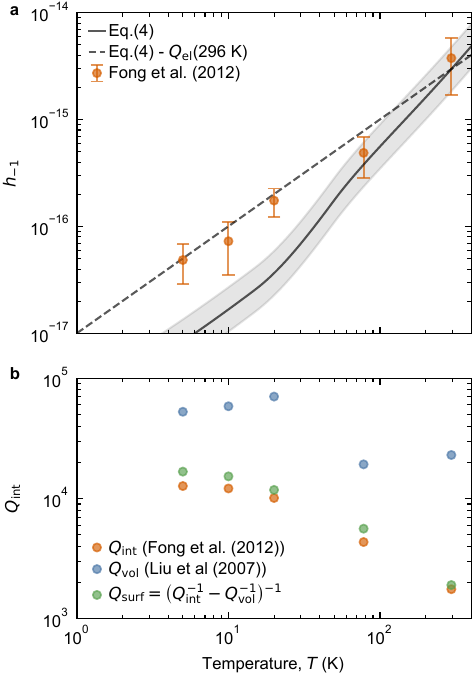}
\caption{Temperature dependence of the flicker floor for the silicon nitride string of Fong \emph{et al.} \cite{Fong2012}. \textbf{a}, Flicker coefficient $\hmo=S_0(T)/f_r^2$, with $S_0$ extracted from Fig.~4 and $f_r=641.6$~kHz from Fig.~1c of Ref.~\cite{Fong2012}. The solid line shows the parameter-free prediction of Eq.~\eqref{eq:main} (no fit), using the measured $Q_\mathrm{int}(T)$ decomposed into volume and surface contributions (Methods), with the surface term weighted for the in-plane mode, and $\sigma_0(T)$ determined from the differential thermal expansion of the film and substrate. The shaded band denotes the uncertainty in $Q_\vol$ and $\beta$. The shaded region denotes the $95\%$ scatter band (see Section~\ref{SI-sec:Qfit}). \textbf{b}, Intrinsic quality factor $Q_\mathrm{int}=Q/D$, obtained from the measured $Q(T)$ (Fig.~1b of Ref.~\cite{Fong2012}) using the dilution factor $D$ of the in-plane mode (Methods), plotted together with the volume quality factor of SiN (Ref.~\cite{liu2007elastic}), and the calculated surface quality factor.}
\label{fig:h-vs-T}
\end{figure}

At a fixed fluctuator population, the relation predicts $\hmo\propto T\,\phi(T)/\sigma_0^2(T)$ [Eq.~\eqref{eq:fdt}]. The only published data set in which the flicker floor and the loss of one silicon nitride resonator were measured over a wide temperature range is that of Fong \emph{et al.} \cite{Fong2012}, who found $\hmo\propto T^{0.94\pm0.10}$ between 5 and 296~K for a stoichiometric string ($380~\mu$m long, $330$~nm thick, $200$~nm wide, in-plane fundamental mode). Fig.~\ref{fig:h-vs-T}(a) compares the measured noise floor with Eq.~\eqref{eq:main} evaluated in two ways: with the elongation loss $Q_\el(T)$ taken from the measured loss at each temperature (solid line), and with $Q_\el$ fixed at its room-temperature value, so that $\hmo\propto T/\sigma_0^2(T)$ (dashed line); the prestress $\sigma_0(T)$ follows from the differential thermal expansion of film and substrate (Methods). Both lines lie within the error bars at 296 and 78~K. Below 78~K, they separate; the data stay within a factor of two of the dashed line, i.e., the loss angle probed by the noise remains close to its room-temperature value while the loss angle measured at $\omega_0$ falls by nearly an order of magnitude. The dashed line is a reference here, not a prediction: it marks what the floor would be if the loss angle in the Allan band did not change. The decomposition of the measured loss in Fig.~\ref{fig:h-vs-T}(b), with the volume loss of LPCVD silicon nitride from Ref.~\cite{liu2007elastic}, suggests that the loss of this device is dominated by its surfaces at all temperatures and that the decrease of the loss on cooling is a decrease of the surface loss. The surface fluctuators thus stop absorbing at $\omega_0$ while they continue to produce noise in the low-frequency Allan band.

The separation of the two lines marks the limit of the structural-damping assumption. Eq.~\eqref{eq:fdt} itself is unaffected, since it requires the loss angle at the frequency of the noise; what no longer holds below 78~K is the identification of that loss angle with the one measured at $\omega_0$. For thermally activated TLS, cooling moves both windows of Eq.~\eqref{eq:window} to lower activation energies, so the temperature series probes the TLS distribution between them. The agreement at 296 and 78~K indicates that the distribution is flat across this interval at these temperatures. Below 78~K, the loss at $\omega_0$ decreases while the noise does not, which could reflect a barrier distribution that thins out at low energies or a crossover from thermally activated to tunnelling relaxation \cite{Phillips1987,seoanez2008surface}. Which of the two applies cannot be decided from the given data. At room temperature, where the floors of all our devices are evaluated, the two loss angles coincide.

\subsection*{Bending-dominated beams \& cantilevers}

Any low-stress film, such as crystalline silicon, lies outside the stress-induced dilution regime. For a doubly clamped beam, the fractional frequency change can be modeled as $\delta(\omega^2)/\omega^2=\delta\sigma_\eff/(\sigma_0+\sigma_c)$ with $\sigma_c=Eh^2\kappa^2/12$ the tensile stress at which the tensile and bending restoring energies of the mode would be equal, i.e. the bending stiffness expressed as a tension, where $\kappa^2=\int w''^2/\int w'^2$ ($\kappa^2L^2=40.7$ for the clamped-clamped fundamental, so $\sigma_c=3.4E(h/L)^2$). The flicker noise model (\ref{eq:main}) carries over with $\sigma_0\to\sigma_0+\sigma_c$ (see Section~\ref{SI-sec:beams}). An unstressed Si beam of $h=100$~nm and $L=10$~$\mu$m has $\sigma_c=57$~MPa and is therefore $(\sigma_0/\sigma_c)^2\approx300$ times more susceptible per fluctuator than a 1~GPa string, with a predicted $(L/h)^4$ dependence of $\hmo V$ at fixed population. A cantilever, having no axial constraint, responds only through the modulus change between the two fluctuator states; this channel is stress-independent and produces no relaxation loss in the linear regime, so for cantilevers, the noise is expected to be decoupled from loss.

This reconciles the surveys \cite{Sansa2016,Roukes2026} with stress dilution. In the model, the invariant of any device family is $\hmo V\propto\phi/(\sigma_0+\sigma_c)^2$: the floor scales with the loss angle of the material and inversely with the square of the total stress, prestress plus bending-equivalent stress. Stressed strings and unstressed beams can therefore lie on one invariant only if the unstressed devices have a correspondingly smaller loss angle. For a 1~GPa string against an unstressed beam with $\sigma_c$ of a few tens of MPa, the required ratio is $(\sigma_c/\sigma_0)^2\sim10^{-2}$ to $10^{-3}$. The unstressed devices in the surveys are largely crystalline, and crystalline resonators exhibit loss angles that are orders of magnitude below those of amorphous films \cite{beccari2022strained,romero2020engineering,hochreiter2025monolithic}, so the required ratio has the expected sign and magnitude.

\subsection*{Conclusion}

Our results indicate that the flicker floor of a tensioned nanomechanical resonator is the thermal noise of its tension. In this picture, the fluctuation-dissipation theorem fixes the tension noise through the loss angle of the material, structural damping gives it the $1/f$ spectrum, and the response of the frequency to the tension dilutes it by the prestress, leading to $\hmo V\sigma_0^2=E\kB T/(2\pi Q_\el)$, with $Q_\el$ the elongation loss obtained from the intrinsic quality factor measured in bending. A microscopic model of two-level systems carrying an eigenstrain reproduces this relation. Silicon nitride resonators follow the relation in volume, stress, thickness and absolute level with one set of independently measured parameters, and the thickness series suggests that noise-active TLS reside both in the volume and at the hydroxyl-terminated surface that limits the quality factor of thin films. 
Published cryogenic data indicate the range of the structural-damping assumption that links the millihertz and megahertz bands. The two bands sample TLS whose activation energies differ by $\kB T\ln(\omega_0/2\pi f)$, a logarithm of a frequency ratio rather than a material property; they are coupled or not according to whether the TLS distribution is flat across that interval. It appears to be down to 78~K, whereas at lower temperatures the surface loss measured at the resonance freezes out while the loss angle probed by the noise remains close to its room-temperature value. At room temperature, the relation implies that the floor of a resonant sensor is lowered by raising the prestress or lowering the intrinsic loss angle of the material, rather than by raising the quality factor through the removal of external loss. It thus offers an account of the gap between the flicker floor and the thermomechanical limit identified a decade ago \cite{Sansa2016} that does not require a new noise source.
\section*{Methods}

\subsection*{Devices and material parameters}
The experiments were performed with strings and square membranes from previous work (Table~\ref{tab:samples}), made from silicon-rich and stoichiometric low-pressure chemical vapor-deposited (LPCVD) silicon nitride on a silicon substrate. The prestress of each device was extracted from its eigenfrequency. The plotting of the parameter-free model curves was done with the material parameters listed in Table~\ref{tab:params} and a device-to-device scattering band of $\times\div 1.58$.

\emph{Dissipation dilution.} In a tensioned resonator the intrinsic loss of the material acts only through the bending energy of the mode, whereas the restoring force is supplied predominantly by the tensile energy. The measured quality factor is therefore enhanced over the intrinsic one by the dilution factor $D$,
\begin{equation}
Q=D\,Q_\mathrm{int},\qquad D=\frac{W_\tens+W_\bend}{W_\bend},
\label{eq:D}
\end{equation}
where $Q_\mathrm{int}^{-1}=\phi_\bend$ is the loss angle of the material under the bending strain of the mode \cite{Schmid2023}. For the $n$-th out-of-plane mode of a string of length $L$ and thickness $h$ with clamped ends \cite{yu2012control},
\begin{equation}
D_n=\left[2\lambda+(n\pi\lambda)^2\right]^{-1},\qquad \lambda=\frac{h}{L}\sqrt{\frac{E}{12\sigma_0}},
\end{equation}
where the first term accounts for the bending near the clamps and the second for the distributed bending of the mode shape \cite{schmid2011damping,Schmid2023}; for an in-plane mode, $h$ is replaced by the width. The corresponding expression considering the biaxial stress for square membranes is taken from Ref.~\cite{Schmid2023}. For each device, $D$ was evaluated with the prestress extracted from its eigenfrequency and the material parameters of Table~\ref{tab:params}, and $Q_\mathrm{int}=Q/D$ was obtained from the measured $Q$. Since external losses such as radiation into the substrate lower $Q$ without changing $D$, the values of $Q_\mathrm{int}$ obtained in this way are lower bounds; the intrinsic loss of a film is given by the upper envelope of a device series.

\begin{table*}[t]
\caption{Inventory of the resonators analyzed in this work. $h$: film thickness; $w$ and $L$: string width and length, or membrane side length $a$; Si/N: film stoichiometry; $\sigma_0$: prestress extracted from the eigenfrequency of each device (range over the group); $\#$: number of devices; Ref.: References of paper where the samples have been presented.}
\label{tab:samples}
\begin{ruledtabular}
\begin{tabular}{lcccccccl}
Group & Geometry & $h$ (nm) & $w$ or $a$ ($\mu$m) & $L$ ($\mu$m) & Si/N & $\sigma_0$ (MPa) & $\#$ &  Ref. \\
\colrule
S56   & string   & 56  & 6     & 100-2000  & 0.89 & 380  & 24    & \cite{sadeghi2019influence} \\
S157  & string   & 157 & 6     & 110       & 0.83 & 785  & 5     & \cite{schmid2011damping} \\
S177  & string   & 177 & 6     & 140-210   & 0.96 & 174  & 4     & \cite{schmid2011damping} \\
S312  & string   & 312 & 6-30     & 210       & 0.84 & 835  & 15     & \cite{schmid2011damping} \\
S340  & string   & 340 & 6-30     & 110       & 0.98 & 200  & 13     & \cite{schmid2011damping} \\
M50   & membrane & 50  &       & 66-2000   & Si-rich   & 160  & 19    & \cite{kanellopulos2025comparative} \\
M30LS   & membrane & 30  &       & 850        & Si-rich   & 134 & 5   & \cite{Cavalleri2026} \\
M30HS   & membrane & 30  &       & 850        & stoich.   & 1070 & 3   & \cite{Cavalleri2026}\\
M300   & membrane & 300  &       &    510    & Si-rich   & 165 & 3   & \\
\end{tabular}
\end{ruledtabular}
\end{table*}

\subsection*{Frequency-noise measurement}
Devices were mounted in a vacuum chamber at $p<10^{-5}$~mbar and read out optically using a laser-Doppler vibrometer (MSA-500 from Polytec), with the fundamental-mode driven piezoelectrically below the Duffing critical amplitude. The resonance frequency was tracked with a phase-locked loop (MFLI from Zurich Instruments), and the Allan deviation $\sigma_y(\tau)$ was computed from the frequency record for over one minute [Fig.~\ref{fig:setup}]. The flicker coefficient was obtained from a joint fit of the additive ($\tau^{-1/2}$), flicker (constant) and drift ($\tau$) branches. 

\emph{Allan variance.} The fractional frequency $y(t)=\delta f(t)/f_0$ is characterized by its one-sided power spectral density $S_y(f)$. The Allan variance at integration time $\tau$ is defined from the averages $\bar y_k$ of $y$ over consecutive intervals of length $\tau$ as $\sigma_y^2(\tau)=\tfrac12\langle(\bar y_{k+1}-\bar y_k)^2\rangle$, and is related to the spectrum by \cite{Schmid2023}
\begin{equation}
\sigma_y^2(\tau)=2\int_0^\infty S_y(f)\,\frac{\sin^4(\pi f\tau)}{(\pi f\tau)^2}\,\dd f .
\label{eq:allan}
\end{equation}
The filter function peaks at $f\approx0.37/\tau$, so that $\sigma_y(\tau)$ samples the frequency noise at Fourier frequencies of order $1/(2\tau)$. For flicker frequency noise, $S_y=\hmo/f$, Eq.~\eqref{eq:allan} gives $\sigma_y^2=2\ln2\,\hmo$, independent of $\tau$; white frequency noise, $S_y=h_0$, gives $\sigma_y^2=h_0/(2\tau)$, and a linear drift $D$ gives $\sigma_y^2=D^2\tau^2/2$.

\subsection*{Intrinsic loss}
Quality factors were measured by ringdown for the fundamental mode per device and the intrinsic loss $Q_\mathrm{int}$ was obtained from the dissipation-dilution relation \cite{Schmid2023,Benga2026}, considering the biaxial strain correction for membranes. The room temperature values of $Q\vol=2.3(3)\times10^{4}$ and $\beta=72(6)$~nm$^{-1}$ ($Q_\vol/(3\beta)$) were extracted by fitting published intrinsic quality factor values \cite{Villanueva2014,Cavalleri2026}. To account for scattering due to inter-chip variability, we use a scattering factor of $\times\div 1.58$ as the uncertainty for $Q_\el$.

\begin{table}[t]
\caption{Material and model parameters used for all model curves and predictions.}
\label{tab:params}
\begin{ruledtabular}
\begin{tabular}{llll}
Symbol & Quantity & Value & Source \\
\colrule
$E$        & Young's modulus              & 200~GPa (on Si)     & \cite{kanellopulos2024stress,klass2022determining} \\
           &                              & 250~GPa (on SiO$_2$) & \cite{klass2022determining} \\
$\nu$      & Poisson's ratio              & 0.23                & \cite{edwards2004comparison} \\
$\rho$     & mass density                 & 3000~kg\,m$^{-3}$   & \cite{Giesriegl2026} \\
$Q_\vol$   & volume-loss $Q$   & $2.3(3)\times10^{4}$ & \ref{SI-sec:Qfit} \\
$\beta$    & surface-loss parameter       & $72(6)$~nm$^{-1}$ & \ref{SI-sec:Qfit} \\
\end{tabular}
\end{ruledtabular}
\end{table}

\subsection*{Data analysis for cryogenic temperature data}
The device studied in Ref.~\cite{Fong2012} is a string resonator of length
$L = 380\,\mathrm{\mu m}$, width $w = 200\,\mathrm{nm}$ and thickness $h = 330\,\mathrm{nm}$,
patterned from a SiN film on a $3\,\mathrm{\mu m}$ thick buried oxide layer on a Si
wafer~\cite{fong2010high}; the measured mode is the fundamental in-plane flexural mode.
To compare the reported noise data with the present measurements, the prestress $\sigma_0(T)$
was first determined at each measurement temperature, since it enters both the flicker
coefficient $h_{-1}(T) = S_0(T)/f_r^2(T)$ and the dilution factor $D$
(Fig.~\ref{fig:h-vs-T}a and b, respectively).

Ref.~\cite{Fong2012} reports the resonance frequency only at $5\,\mathrm{K}$,
$f_r(5\,\mathrm{K}) = 641.6\,\mathrm{kHz}$ (Fig.~1c of Ref.~\cite{Fong2012}). The string relation
$f_r = (2L)^{-1}\sqrt{\sigma_0/\rho}$, with $\rho = 3000\,\mathrm{kg\,m^{-3}}$, yields
$\sigma_0(5\,\mathrm{K}) = 713\,\mathrm{MPa}$. Since the ends of the string follow the frame,
whose thermal expansion is dominated by the Si substrate, the prestress at higher temperatures
is set by the differential thermal expansion of substrate and film,
\begin{equation}
\sigma_0(T) = \sigma_0(5\,\mathrm{K}) + E \int_{5\,\mathrm{K}}^{T}
\left[ \alpha_{\mathrm{Si}}(T') - \alpha_{\mathrm{SiN}}(T') \right] \mathrm{d}T',
\label{eq:sigma_T}
\end{equation}
where $\alpha_{\mathrm{Si}}$ and $\alpha_{\mathrm{SiN}}$ were taken from
Refs.~\cite{middelmann2015thermal} and ~\cite{Chuang2005}, respectively and $E = 250\,\mathrm{GPa}$, in line with values reported for SiN thin
films on silica substrates~\cite{klass2022determining}.
The temperature dependence of $E$ (about $2\%$ between
$30$ and $298\,\mathrm{K}$~\cite{Chuang2005}) and of $\rho$ (about $0.1\%$) was neglected.
The resonance frequency was then extracted for each temperature as $f_r(T) = f_r(5\,\mathrm{K})\sqrt{\sigma_0(T)/\sigma_0(5\,\mathrm{K})}$.

To evaluate $Q_\mathrm{el}(T)$ in Eq.~\eqref{eq:main} (solid line in Fig.~\ref{fig:h-vs-T}a),
the volume loss $Q_\mathrm{vol}^{-1}(T)$ of the SiN film was taken from the internal friction of
as-deposited LPCVD SiN films measured with double-paddle oscillators~\cite{liu2007elastic},
a geometry in which the film strain is nearly uniform across the thickness and volume loss
dominates. [Above approximately $100\,\mathrm{K}$, where the oscillator background dominates
those measurements, $Q_\mathrm{vol}$ was set to the value obtained in the present work,
$Q_\mathrm{vol} = 2.3\times 10^{4}$, which the data of Ref.~\cite{liu2007elastic} approach.]
The surface contribution then follows from the measured intrinsic loss as
\begin{equation}
Q_\mathrm{surf}^{-1}(T) = Q_\mathrm{int}^{-1}(T) - Q_\mathrm{vol}^{-1}(T)
= D(T)\,Q_\mathrm{exp}^{-1}(T) - Q_\mathrm{vol}^{-1}(T).
\label{eq:Qsurf}
\end{equation}
For the in-plane mode, the surface term in $Q_\mathrm{el}$ is weighted by
$m = (3h + w)/(h + w) \approx 2.25$ instead of $3$, since the sidewalls are located at the maximum
bending strain while the top and bottom faces experience the mean strain of the cross-section.


\bibliographystyle{apsrev4-2}
\bibliography{refs}

\section*{Acknowledgements}
The authors thank Pedram Sadeghi for fabricating the tapered strings. We thank Guillermo Arregui Bravo and Alessio Zicoschi for the discussions. The supporting derivations in \ref{SI-sec:micro-model}, \ref{SI-sec:relations}, and \ref{SI-sec:beams} of the supplementary information were developed using Claude Fable 5.1 from Anthropic; all steps were verified by the authors.

\section*{Funding}
This project received funding from the Austrian Science Fund (FWF) under project 10.55776/I6086 and the Novo Nordisk Foundation under grant no. NNF22OC0077964 - MASMONADE.

\end{document}


\title{Supplementary Information:\\
1/f frequency-noise reduction in nanomechanical resonators by dissipation dilution}

\author{Kostas Kanellopulos}
\affiliation{Institute of Sensor and Actuator Systems, TU Wien, Gusshausstrasse 27-29, 1040 Vienna, Austria}
\author{Silvan Schmid}
\email{silvan.schmid@tuwien.ac.at}
\affiliation{Institute of Sensor and Actuator Systems, TU Wien, Gusshausstrasse 27-29, 1040 Vienna, Austria}

\maketitle

\tableofcontents

\section{Fit of the intrinsic quality factor}
\label{sec:Qfit}

Values of $Q_\mathrm{intr}$ for SiN resonators of thickness $h$ were taken from
the meta-analysis of Ref.~\cite{Villanueva2014} (22 data points, digitized from
the published figure, excluding the data point marked as uncertain there) and
from the as-received membranes measured in high vacuum in
Ref.~\cite{Cavalleri2026} (5 data points at $h = 30$ and $50$~nm, extracted
from the vector graphics of the published figure). The 10~nm membranes of
Ref.~\cite{Cavalleri2026} were excluded, as their dissipation is dominated by
an additional excess-loss channel of the defective nucleation layer that is
not captured by the two-channel model. The combined data set ($N = 27$) was
fitted with
\begin{equation}
Q_\mathrm{intr} = \left[(\beta h)^{-1} + Q_\vol^{-1}\right]^{-1}
\label{eq:Qintr}
\end{equation}
by unweighted nonlinear least squares on $\ln Q_\mathrm{intr}$, since the
sample-to-sample scatter is multiplicative. The standard deviation of the
logarithmic residuals corresponds to a scatter factor of 1.26 (0.10 decades),
which was used to scale the parameter covariance matrix. The fit yields
$Q_\vol = 2.3(3)\times10^{4}$ and $\beta = 72(6)~\mathrm{nm}^{-1}$
(1$\sigma$), with a correlation coefficient of $-0.57$.
The uncertainty of $h^{*} = Q_\vol/(3\beta) = 107(21)$~nm was obtained by
linear propagation including this correlation and agrees with a nonparametric
bootstrap (5000 resamples, 68\% interval 88 to 127~nm). Since the data points
originate from a limited number of wafers and laboratories and are therefore
not fully independent, the quoted uncertainties are statistical lower bounds.

\begin{figure}
\centering
\includegraphics[width=\columnwidth]{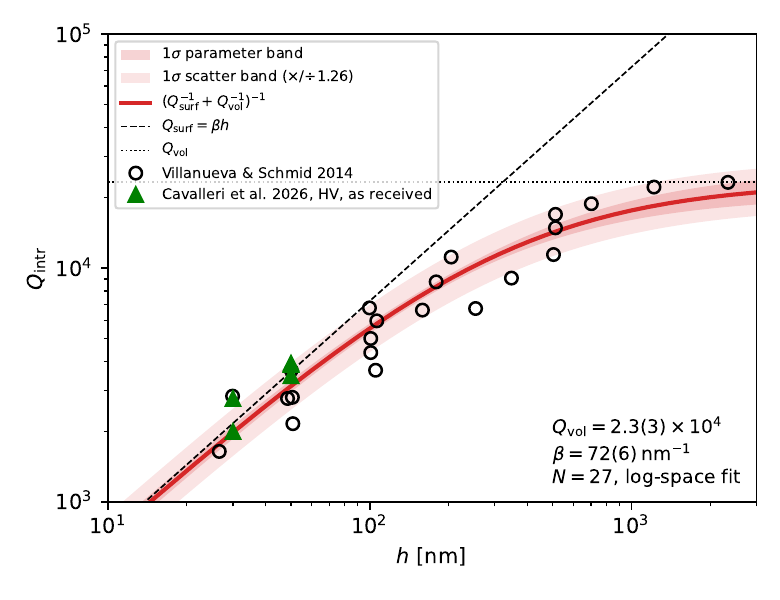}
\caption{Intrinsic quality factor of SiN resonators as a function of film
thickness $h$. Open circles: meta-analysis data of Ref.~\cite{Villanueva2014};
triangles: as-received membranes measured in high vacuum,
Ref.~\cite{Cavalleri2026}. The solid line is a fit of Eq.~\eqref{eq:Qintr}
to the combined data ($N = 27$), yielding $Q_\vol = 2.3(3)\times10^{4}$
and $\beta = 72(6)~\mathrm{nm}^{-1}$. The dashed and dotted lines indicate the
surface-loss limit $Q_\mathrm{surf} = \beta h$ and the volume-loss limit
$Q_\vol$, respectively. The dark shaded band shows the $1\sigma$
uncertainty of the fit, propagated from the parameter covariance; the light
band shows the $1\sigma$ sample-to-sample scatter of the data (factor 1.26).}
\label{fig:Qintr-fit}
\end{figure}

\section{Stress dilution versus stoichiometry}
\label{sec:stoich}

\begin{figure}
\centering
\IfFileExists{figS2_stoichiometry.pdf}{%
  \includegraphics[width=\columnwidth]{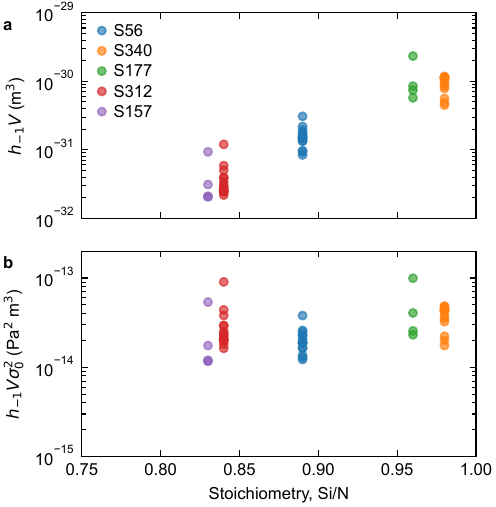}}{%
  \fbox{\parbox[c][4cm][c]{0.9\columnwidth}{\centering fig4\_stoichiometry.pdf}}}
\caption{Stress dilution against stoichiometry. (a) $\hmo V$ and
(b) $\hmo V\sigma_0^2$ against the Si/N ratio of the film for the devices of
Fig.~\ref{main-fig:stress} of the main text, grouped by thickness. The
prestress decreases with increasing silicon content, so (a) mirrors the stress
dilution of Fig.~\ref{main-fig:stress}; (b) is flat within the device-to-device
scatter, at the value predicted by the volume loss.}
\label{fig:stoich}
\end{figure}

In LPCVD silicon nitride, the prestress is set by the composition: silicon-rich
films are low-stress, stoichiometric films are high-stress. The $\sigma_0^{-2}$
trend of Fig.~\ref{main-fig:stress} of the main text could therefore be a
composition dependence of the fluctuator density $K_v$ in disguise; plotted
against the Si/N ratio, $\hmo V$ rises by a factor ${\sim}30$ from
Si/N $=0.83$ to $0.98$ [Fig.~\ref{fig:stoich}(a)]. The loss excludes this
reading. In the model the same $K_v$ that sets the noise sets the volume loss,
$Q_\vol^{-1}\propto K_v/\ln R$ [Eq.~\eqref{eq:vol}], an
identification that the absolute value of the floor confirms; a factor
30 in $K_v$ across the stoichiometry range would therefore appear as a
factor 30 in the intrinsic loss at fixed thickness. Instead, all devices
lie on one $Q_\mathrm{intr}(h)$ curve irrespective of composition
[Fig.~\ref{fig:Qintr-fit}],
as found earlier for silicon nitride of both compositions
\cite{Villanueva2014}. The loss-active ensemble is therefore the same in all
films, and with $K_v$ fixed, the noise can vary only through $\sigma_0$.
Accordingly, $\hmo V\sigma_0^2$ is flat across the stoichiometry range at
around $(2$ to $3)\times10^{-14}$~Pa$^2$\,m$^3$ [Fig.~\ref{fig:stoich}(b)],
while $\hmo V$ varies by a factor 30. Two composition-independent checks
agree. With a fixed composition (Si/N $=0.83$ to $0.84$), the 157 and 312~nm
devices share one $\hmo V Q_\el$ [Fig.~\ref{main-fig:stress}].
A fluctuator density that tracked $\sigma_0^2$ exactly, with one prefactor
across films from different furnaces and vendors, is not a credible
alternative.

\section{Microscopic model: two-state fluctuators}
\label{sec:micro-model}

The FDT fixes the relation between tension noise and loss without saying what the fluctuators are. A microscopic model is needed for two things: to show that the loss angle is the same at millihertz and at megahertz, and to say how fluctuators at the surfaces and in the volume enter. It is also where the two pictures of the floor in the literature, modulus fluctuators and strain fluctuators, can be compared on equal terms. This section is written to be followed without prior knowledge of two-level-system physics; the detailed integrals are in Section~\ref{sec:relations}. 

\subsection{A two-state fluctuator}
\begin{figure}
\centering
\includegraphics[width=\columnwidth]{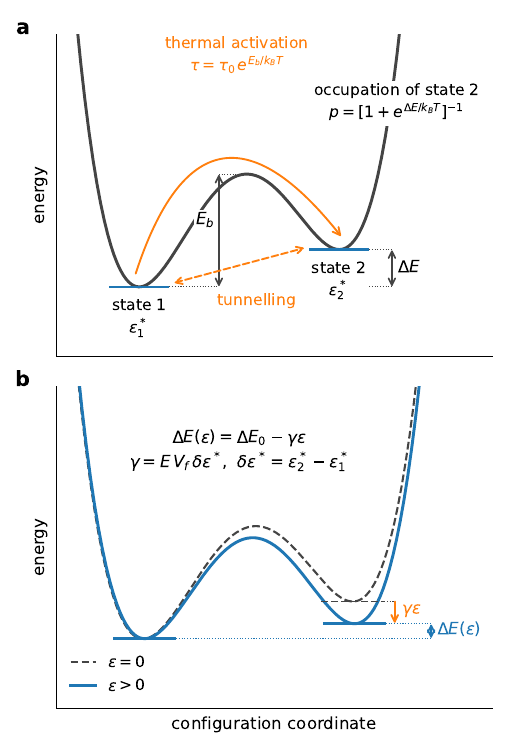}
\caption{Two-level system (TLS) and its coupling to strain. (a) Double-well potential of a defect that can occupy two atomic configurations with stress-free strains $\varepsilon^*_1$ and $\varepsilon^*_2$, separated by the barrier $E_b$ and differing in energy by $\Delta E$. The occupation of the upper state is $p=[1+e^{\Delta E/\kB T}]^{-1}$. Switching proceeds by thermal activation over the barrier, with correlation time $\tau=\tau_0\,e^{E_b/\kB T}$, or by tunnelling through it. (b) An applied strain $\varepsilon$ tilts the potential and changes the splitting by $-\gamma\varepsilon$, with the deformation potential $\gamma=EV_f\,\delta\varepsilon^*$, Eq.~\eqref{eq:gamma}; this linear coupling produces both the frequency noise and the relaxation loss.}
\label{fig:TLS}
\end{figure}

A two-level system (TLS) is a defect that can occupy one of two atomic configurations separated by an energy barrier; it switches between them by thermal activation over the barrier at high temperature and by tunneling through it at low temperature \cite{Phillips1987} [see schematic in Fig.~\ref{fig:TLS}(a)]. Three quantities characterize it. The \emph{energy splitting} $\Delta E$ is the energy difference between the two configurations; it sets the thermal occupation of the upper state, $p=[1+\exp(\Delta E/\kB T)]^{-1}$, which is $1/2$ for a symmetric defect and small for a strongly asymmetric one. The \emph{activation energy} $E_b$ is the barrier between the configurations; it sets the \emph{correlation time} $\tau=\tau_0\exp(E_b/\kB T)$ with which the defect switches back and forth, with $\tau_0\sim10^{-13}$~to~$10^{-12}$~s an attempt time. A defect switching with correlation time $\tau$ produces a random telegraph signal whose spectrum is a Lorentzian of width $1/\tau$. Defects of this kind are the standard description of the low-temperature properties of glasses \cite{Phillips1987}, and they have recently been observed one at a time in nanomechanical resonators \cite{Maksymowych2025Frequency}.

\subsection{How a fluctuator couples to strain}
\label{sec:coupling}
The coupling follows from the elastic energy stored in the volume $V_f$ of the defect. Treat $V_f$ as an inclusion whose stress-free state is strained by $\varepsilon^*$ relative to the surrounding material, and let the resonator impose a strain $\varepsilon$ at its location. The inclusion is then elastically strained by $\varepsilon-\varepsilon^*$. For a string, where the stress is uniaxial, its elastic energy is
\begin{equation}
\begin{split}
W_f(\varepsilon)&=\tfrac12EV_f(\varepsilon-\varepsilon^*)^2\\
&=\underbrace{\tfrac12EV_f\varepsilon^2}_{\text{elastic energy}}
\;\underbrace{-\;EV_f\varepsilon^*\varepsilon}_{\text{interaction}}
\;+\;\underbrace{\tfrac12EV_f\varepsilon^{*2}}_{\text{self-energy}} .
\label{eq:Wf}
\end{split}
\end{equation}
The two configurations of a defect can differ in $\varepsilon^*$ or in $E$, and these are the two kinds of fluctuator.

\emph{Eigenstrain fluctuator.} The configurations have the same stiffness but different stress-free strains, $\varepsilon^*_1$ and $\varepsilon^*_2$; a reorienting elastic dipole is the standard example. The elastic energy in Eq.~\eqref{eq:Wf} is then the same in both states, the self-energy differs but does not depend on $\varepsilon$ and is absorbed in the asymmetry $\Delta E_0$, and the interaction term is linear in $\varepsilon$ [Fig.~\ref{fig:TLS}(b)]. The energy splitting is
\begin{equation}
\Delta E(\varepsilon)=\Delta E_0-\gamma\,\varepsilon,\quad \gamma\equiv EV_f\,\delta\varepsilon^*,\;\delta\varepsilon^*=\varepsilon^*_2-\varepsilon^*_1 ,
\label{eq:gamma}
\end{equation}
with $\gamma$ the deformation potential. The interaction term is exact for an inclusion of any shape: by Eshelby's theorem the interaction energy of an inclusion with an applied stress $\sigma^A_{ij}$ is $-\int_{V_f}\sigma^A_{ij}\varepsilon^*_{ij}\,\dd V$, while the self-energy depends on shape but not on the applied strain. The general form also shows how the geometry enters: for a string only $\sigma_{xx}=E\varepsilon$ is non-zero, so $\gamma=EV_f\varepsilon^*_{xx}$; for a membrane under equibiaxial strain, $\sigma_{xx}=\sigma_{yy}=E\varepsilon/(1-\nu)$, and the biaxial modulus replaces $E$ in both the noise and the loss, so that the relation between them is unchanged to within a few percent (Section~\ref{sec:geometry}).

\emph{Modulus fluctuator.} The configurations have the same stress-free strain but different stiffness, $E$ and $E+\delta E_f$. Now the interaction and self-energy terms cancel in the difference between the states and the elastic term survives,
\begin{equation}
\Delta E(\varepsilon)=\Delta E_0-\tfrac12V_f\,\delta E_f\,\varepsilon^2 ,
\label{eq:modulus}
\end{equation}
quadratic in strain, with no linear term at $\varepsilon=0$. This is the coupling assumed in Refs.~\cite{Cleland2002,Roukes2026}. The two couplings are the two ways in which Eq.~\eqref{eq:Wf} can depend on the state of the defect: through $\varepsilon^*$, linearly, or through $E$, quadratically. We treat the eigenstrain fluctuator first and return to the modulus fluctuator in Section~\ref{sec:modulus}.

\subsection{Many fluctuators: the distribution of correlation times}
\label{sec:band}
A resonator contains $N$ independent fluctuators with a broad spread of activation energies, hence of correlation times. We assume that the correlation times are distributed log-uniformly over a band $[\tau_{\min},\tau_{\max}]$, with $R=\tau_{\max}/\tau_{\min}$, independently of $\gamma$ and $\Delta E_0$. For thermally activated switching this follows from a flat distribution of activation energies of width $\Delta E_b$, since $\dd\ln\tau=\dd E_b/\kB T$, and then $\ln R=\Delta E_b/\kB T$. The fraction of fluctuators with correlation times within a factor $e$ of a given value is $1/\ln R$.

Two consequences of this distribution are used below. A sum of Lorentzians with log-uniformly distributed widths is a $1/f$ spectrum within the band (the Dutta--Horn mechanism, \cite{Dutta1981}): at Fourier frequency $f$ the fluctuators with $1/\tau\sim2\pi f$ dominate, and their number per unit $\ln\tau$ is the same at every $f$. By the same token, the relaxation loss at a mode frequency $\omega_0$ inside the band is independent of $\omega_0$; this is structural damping. The band must contain both $1/\omega_0$ and the integration times $\tau$ over which the Allan plateau is observed; this is the microscopic form of the structural-damping hypothesis. In terms of activation energies, a measurement at frequency $\omega$ and temperature $T$ probes the fluctuators with $E_b\approx\kB T\ln(1/\omega\tau_0)$: about 0.4~eV for the loss at a few megahertz and 0.7 to 0.9~eV for the noise at 0.01 to 1~Hz at room temperature. The relation between noise and loss therefore requires the distribution of activation energies to be flat between these two energies, which is the usual assumption for the broad distributions of amorphous solids but is not guaranteed, and which the decoupling argument of Ref.~\cite{Roukes2026} questions. The data decide; at room temperature they support it, at cryogenic temperatures (main text, temperature dependence) the two energies become small enough that the assumption is less secure.

\subsection{Frequency noise of eigenstrain fluctuators}
When fluctuator $i$ switches, its stress-free strain changes by $\delta\varepsilon^*_i$ over $V_{f,i}$. The ends of the resonator are fixed, so the total elongation cannot change; the elastic strain of the rest of the resonator must compensate, and the mean tension changes by
\begin{equation}
\delta\sigma_i=-\frac{\gamma_i}{V}
\label{eq:dsigma}
\end{equation}
(derivation in Section~\ref{sec:shift}). This stress change is uniform along the resonator, because force balance makes the tension uniform, and it does not depend on where in the cross section the fluctuator sits. The resonance frequency follows the tension, $\delta\omega/\omega=\delta\sigma/2\sigma_0$, so the frequency switches between two values separated by
\begin{equation}
\Delta_i=\frac{\gamma_i}{2\sigma_0V}.
\label{eq:Delta}
\end{equation}
The prestress appears in the denominator: the same stress jump is a smaller fractional change of a larger tension. This is the origin of the stress dilution of the floor. The fluctuator also produces a local, self-equilibrated stress field around itself, but that field does not change the mode-weighted tension to any relevant order (Section~\ref{sec:shift}), so Eq.~\eqref{eq:Delta} holds for any fluctuator position and any resonator width.

Summing the Lorentzian spectra of the $N$ switching processes over the log-uniform band gives $S_y(f)=\hmo/f$ with
\begin{equation}
\hmo=\frac{N\,\langle\gamma^2p(1-p)\rangle}{4\,\sigma_0^2V^2\ln R},
\label{eq:hm1}
\end{equation}
and a flat Allan deviation, $\sigma_y^2=2\ln2\,\hmo$. Each fluctuator contributes the square of its frequency jump times its variance $p(1-p)$, divided by $\ln R$ because only the fraction of fluctuators with correlation times near $1/2\pi f$ contributes at each $f$ (Section~\ref{sec:flicker}).

\subsection{Dissipation of the same fluctuators}
\label{sec:dissipation}
The same coupling makes the fluctuators absorb energy from the vibrating mode. Under a slowly varying strain the equilibrium occupation is $p_\mathrm{eq}(\varepsilon)=[1+e^{\Delta E(\varepsilon)/\kB T}]^{-1}$, so a strain change $\delta\varepsilon$ shifts the occupation by
\begin{equation}
\delta p=\frac{\partial p_\mathrm{eq}}{\partial\varepsilon}\,\delta\varepsilon=\frac{\gamma}{\kB T}\,p(1-p)\,\delta\varepsilon .
\end{equation}
The occupation does not follow instantly; it relaxes toward $p_\mathrm{eq}$ with the correlation time $\tau$. Under the oscillating strain of the mode it therefore lags, and the fluctuator, which exerts the generalized force $\gamma p$ on the strain (Eq.~\eqref{eq:gamma}), does net work per cycle. With $\delta p=\chi(\omega)\,\delta\varepsilon$, $\chi=\chi_0/(1+i\omega\tau)$, $\chi_0=\gamma^2p(1-p)/\kB T$, the energy dissipated per cycle at strain amplitude $\varepsilon_a$ is
\begin{equation}
\Delta W=\pi\varepsilon_a^2\,\chi''(\omega)=\pi\varepsilon_a^2\,\frac{\gamma^2p(1-p)}{\kB T}\,\frac{\omega\tau}{1+\omega^2\tau^2},
\label{eq:dW}
\end{equation}
the relaxation absorption of a two-level system. The loss is quadratic in $\gamma$ because it is the product of two linear factors: how strongly the strain drives the occupation, and how strongly the occupation acts back on the strain. Averaged over the log-uniform band, the Debye factor $\omega\tau/(1+\omega^2\tau^2)$ gives the constant $\Lambda=\pi/(2\ln R)$, independent of $\omega$.

Which strain drives the fluctuators? The only dynamic strain of first order in the mode amplitude is the bending strain $\varepsilon_b=-z\,w''(x)$, which varies linearly across the thickness and vanishes on the neutral plane; the elongation strain of a vibrating string is quadratic in the amplitude. The loss therefore samples the fluctuators with the weight $z^2$. Averaged over the thickness this weight is $h^2/12$; at the surfaces it is $h^2/4$, three times larger. This is the one place where the location of a fluctuator enters the loss differently from the noise, in which every fluctuator counts with the same weight (Eq.~\eqref{eq:Delta}).

\subsection{The relations}
Place the ensemble in the volume, $N=n_vV$ with $K_v\equiv n_v\langle\gamma^2p(1-p)\rangle$, or on both faces, $N=2n_sA$ with $K\equiv n_s\langle\gamma^2p(1-p)\rangle$. Normalizing the dissipated energy to the bending energy of the mode (Section~\ref{sec:dissipation-details}) gives
\begin{align}
\text{volume:}\ \hmo V\sigma_0^2&=\frac{K_v}{4\ln R}, & Q_\vol^{-1}&=\frac{\pi K_v}{2E\kB T\ln R},
\label{eq:vol}\\
\text{surface:}\ \hmo V\sigma_0^2&=\frac{K}{2h\ln R}, & Q_\surf^{-1}&=\frac{3\pi K}{Eh\kB T\ln R}.
\label{eq:surf}
\end{align}
Eq.~\eqref{eq:vol} gives a thickness-independent volume loss and Eq.~\eqref{eq:surf} the surface-loss law $Q_\surf=\beta h$ established for silicon nitride \cite{Villanueva2014,Benga2026}, with $\beta=E\kB T\ln R/(3\pi K)$. In either placement the ensemble enters noise and loss only through $K/\ln R$, which cancels in the product:
\begin{equation}
\hmo V\,Q_\vol=\frac{E\kB T}{2\pi\sigma_0^2},\qquad
\hmo V\,Q_\surf=\frac{E\kB T}{6\pi\sigma_0^2}.
\label{eq:mainboth}
\end{equation}
The volume relation is Eq.~\eqref{main-eq:main} of the main text, i.e. the FDT with $\phi=1/Q_\vol$, as it must be for a thermal ensemble: the microscopic model adds no new content to the volume relation, only the conditions under which it holds. The factor 3 of the surface relation is the ratio of the bending weight at the surface to its mean, derived above: for equal numbers of fluctuators the noise is the same, the measured surface loss three times larger. It is what defines the elongation loss $Q_\el$ of Eq.~\eqref{main-eq:main}. The relations do not require that the same individual fluctuators produce the noise at millihertz and the loss at megahertz; they require one distribution whose density in $\ln\tau$ is the same in both bands (Section~\ref{sec:band}).

\subsection{Modulus fluctuators}
\label{sec:modulus}
Now return to the second coupling, Eq.~\eqref{eq:modulus}. Its consequences depend on whether the resonator is prestressed.

\emph{Without prestress.} The coupling is quadratic: a small oscillating strain $\delta\varepsilon=\varepsilon_a\cos\omega t$ gives $\Delta E(t)=\Delta E_0-\tfrac14V_f\delta E_f\varepsilon_a^2(1+\cos2\omega t)$, which modulates the splitting at $2\omega$ but not at $\omega$. Nothing drives the occupation at the mode frequency, so there is no relaxation loss at first order in the mode energy. Switching still changes the stiffness of the defect volume and shifts the frequency of a resonator whose stiffness sets its frequency, a cantilever or an unstressed beam, by $\delta\omega/\omega=V_f\delta E_f/(2EV)$ (Section~\ref{sec:beams}). The modulus fluctuator thus produces frequency noise without producing loss; noise and loss are decoupled, and the conclusion of Ref.~\cite{Roukes2026} holds for such resonators.

\emph{Under prestrain.} Let the resonator carry a prestrain $\varepsilon_0=\sigma_0/E$ and write $\varepsilon=\varepsilon_0+\delta\varepsilon$. Expanding Eq.~\eqref{eq:modulus},
\begin{equation}
\Delta E=\Delta E_0-\tfrac12V_f\delta E_f\varepsilon_0^2-\underbrace{V_f\,\delta E_f\,\varepsilon_0}_{\gamma_\eff}\,\delta\varepsilon-\tfrac12V_f\delta E_f\,\delta\varepsilon^2 .
\label{eq:expand}
\end{equation}
The prestrain generates a term linear in $\delta\varepsilon$. To first order the modulus fluctuator has become an eigenstrain fluctuator with the effective deformation potential $\gamma_\eff=V_f\delta E_f\varepsilon_0$: the stiffer state carries more stress at the same strain, and a change of modulus at fixed prestrain is a change of stress. Everything derived above applies with $\gamma\to\gamma_\eff$. Its frequency jump is
\begin{equation}
\Delta=\frac{\gamma_\eff}{2\sigma_0V}=\frac{V_f\,\delta E_f\,\varepsilon_0}{2\sigma_0V}=\frac{V_f\,\delta E_f}{2EV},
\end{equation}
independent of stress, which is the result of Refs.~\cite{Cleland2002,Roukes2026}; its relaxation loss, Eq.~\eqref{eq:dW}, is proportional to $\gamma_\eff^2\propto\sigma_0^2$. Inserted into Eq.~\eqref{main-eq:fdt}, a loss angle proportional to $\sigma_0^2$ returns a stress-independent floor: the FDT reproduces the modulus-channel result, as it must, since a thermal ensemble cannot have a stress-independent noise together with a stress-independent loss.

\emph{What distinguishes the channels.} The relation holds for both kinds of fluctuator. They differ in how the loss angle depends on prestress:
\begin{center}
\begin{tabular}{@{}lcc@{}}
\hline
channel & $\phi(\sigma_0)$ & $\hmo(\sigma_0)$ \\
\hline
eigenstrain & const. & $\propto\sigma_0^{-2}$ \\
modulus, prestrained & $\propto\sigma_0^{2}$ & const. \\
modulus, unstressed$^{\mathrm a}$ & none$^{\mathrm b}$ & $\delta E/2E$ \\
\hline
\multicolumn{3}{@{}l@{}}{\footnotesize $^{\mathrm a}$cantilever; $^{\mathrm b}$no first-order loss, decoupled from noise}
\end{tabular}
\end{center}
Whether the floor dilutes with stress is therefore decided by how the intrinsic loss depends on stress. If the loss of silicon nitride were dominated by modulus fluctuators, $Q_\mathrm{int}$ would fall as $\sigma_0^{-2}$, by a factor of about 25 between the 200~MPa and 1~GPa films. The intrinsic loss of silicon nitride is independent of composition and hence of prestress (Section~\ref{sec:Qfit}, \cite{Villanueva2014}), so the modulus channel is at most a minor part of $\phi$ and the floor follows the first row. A doubly clamped beam and a cantilever of the same film would measure the ratio of the two couplings directly (Section~\ref{sec:beams}).

\section{Derivation details}\label{sec:relations}
This section supplies the steps behind the results stated in Section~\ref{sec:micro-model}, in the same order: the stress change of a single TLS and why its position does not matter, the sum over the rate distribution that yields the $1/f$ spectrum, the dissipation integrals that fix the surface factor, and the biaxial constraint of a membrane.

\subsection{Frequency shift of one TLS}
\label{sec:shift}
The ends of the resonator are fixed, so the geometric strain $\varepsilon_g$ is prescribed. Let the TLS occupy a slice of length $\ell$ of a string of length $L$, with eigenstrain $\delta\varepsilon^*$ on the slice. Force balance makes the tension uniform along the string, and compatibility, $\varepsilon_g=L^{-1}\int_0^L[\varepsilon_\mathrm{el}+\varepsilon^*(x)]\,\dd x$, gives $\delta\sigma=-E\,\delta\varepsilon^*V_f/V=-\gamma/V$, Eq.~\eqref{eq:dsigma}, undiluted by $\sigma_0/E$. For a TLS smaller than the cross section, the stress perturbation consists of this mean-stress change, enforced by the boundary, and a self-equilibrated local field. The frequency depends on the mode-weighted stress, $\delta(\omega^2)/\omega_0^2=\delta W_\tens/W_\tens$ with $W_\tens=\tfrac12\sigma_0h\int|\nabla\psi|^2\dd A$ ($\psi$ the mode shape). The mean-stress change contributes $\delta\sigma/\sigma_0$ exactly. The local field contributes nothing to leading order. For a string, equilibrium with free lateral edges, $\partial_x\sigma_{xx}+\partial_y\sigma_{xy}=0$, makes the cross-section integral $\int\sigma_{xx}\dd y$ of the local field independent of $x$, and since the field vanishes far from the TLS this integral is zero at every $x$; the mode weight of a string depends on $x$ only, so the local field drops out identically, for any fluctuator position and any string width. For a membrane the exterior field of an inclusion is dipole-like in the plane and integrates to zero over any circle around the TLS, while the mode weight varies on the scale of the membrane; the overlap is of order $(a_f/a)^2$ for a TLS of size $a_f$ in a membrane of side $a$, and is negligible. Hence $\delta\sigma_\eff=\delta\sigma=-\gamma/V$ and Eq.~\eqref{eq:Delta} follows from $\Delta=\tfrac12|\delta\sigma|/\sigma_0$. An off-centre inclusion adds a static bending moment, which changes the frequency only at second order in $\delta\varepsilon^*$. A fluctuator also changes the local bending stiffness; in the dilution regime this is suppressed by $(n\pi\lambda)^2$ with $\lambda=(h/L)\sqrt{E/12\sigma_0}$ and is neglected.

\subsection{Flicker noise}
\label{sec:flicker}
With $s_i(t)\in\{0,1\}$ the state of TLS $i$, $y(t)=\sum_i\Delta_i[s_i(t)-p_i]$. A stationary two-state Markov process with occupation $p$ and correlation time $\tau$ has variance $p(1-p)$ and one-sided spectral density $\Delta^2p(1-p)\,4\tau/[1+(2\pi f\tau)^2]$. With the factorization of the rate distribution,
\begin{align}
S_y(f)&=N\langle\Delta^2p(1-p)\rangle\frac{1}{\ln R}\int\frac{4\tau\,\dd(\ln\tau)}{1+(2\pi f\tau)^2}\notag\\
&=\frac{N\langle\Delta^2p(1-p)\rangle}{f\ln R}
\end{align}
for $f$ inside the band, using $\int_0^\infty\dd u/(1+u^2)=\pi/2$. Inserting Eq.~\eqref{eq:Delta} gives Eq.~\eqref{eq:hm1}, and $N=n_vV$ or $N=2n_sA$ gives the noise parts of Eqs.~\eqref{eq:vol} and \eqref{eq:surf}. The Allan variance of $S_y=\hmo/f$ is $2\hmo\int_0^\infty\sin^4x/x^3\,\dd x=2\ln2\,\hmo$, independent of $\tau$.

\subsection{Dissipation}
\label{sec:dissipation-details}
With $\Delta W$ from Eq.~\eqref{eq:dW} and the bending strain $\varepsilon_b=-zw''(x)$, the volume population of a string with $W_\bend=\tfrac12E(bh^3/12)\int w''^2\dd x$ gives
\begin{equation}
\sum_i\Delta W_i=n_v\pi\,b\frac{h^3}{12}\frac{\langle\gamma^2p(1-p)\rangle}{\kB T}\Lambda\int w''^2\dd x,
\end{equation}
and the surface population, with $\varepsilon_a^2=(h^2/4)w''^2$ on both faces,
\begin{equation}
\sum_i\Delta W_i=2n_s\pi\,b\frac{h^2}{4}\frac{\langle\gamma^2p(1-p)\rangle}{\kB T}\Lambda\int w''^2\dd x,
\end{equation}
where $\Lambda=\langle\omega_0\tau/(1+\omega_0^2\tau^2)\rangle_\tau=\pi/(2\ln R)$ for the log-uniform band with $1/\omega_0$ inside it. Dividing by $2\pi W_\bend$ gives the loss parts of Eqs.~\eqref{eq:vol} and \eqref{eq:surf}; the $\int w''^2$ cancel, and the ratio of the two results for equal numbers of TLS ($n_vh=2n_s$) is $1/3$. In the lossy-layer decomposition $Q_\surf^{-1}=(6\delta/h)Q_S^{-1}$ used in Refs.~\cite{Benga2026,Cavalleri2026}, $\beta=Q_S/(6\delta)$; only this combination is measured. The same layer under uniform elongation stores the fraction $2\delta/h$ of the energy rather than $6\delta/h$, which is the factor $1/3$ once more. Dividing the noise by the loss gives Eq.~\eqref{eq:mainboth}. The factor $\kB T$ in the numerator is the temperature factor of the FDT: in the two-state language it appears because the susceptibility of a thermal two-state system carries $1/\kB T$ (Eq.~\eqref{eq:dW}) while its variance $p(1-p)$ does not.

\subsection{Membranes: the biaxial constraint}
\label{sec:geometry}
A membrane is constrained in both in-plane directions, which changes the coupling of a TLS to the resonator relative to a string. The change enters the noise and the loss through the same factor and cancels in their ratio to within a few percent, so no geometry factor is needed in the relation.

For an isotropic in-plane eigenstrain $\delta\varepsilon^*$ in a plate under plane stress, the interaction energy is $-V_f\delta\varepsilon^*(\sigma_{xx}+\sigma_{yy})$, and since $\sigma_{xx}+\sigma_{yy}=E\,\mathrm{tr}\,\varepsilon/(1-\nu)$, the TLS couples to the trace of the strain with the deformation potential
\begin{equation}
\gamma_m=\frac{\gamma}{1-\nu},\qquad \gamma=EV_f\delta\varepsilon^* ,
\end{equation}
where $\gamma$ is the value for a string, whose transverse strain is free. This one factor enters both observables.

\emph{Noise.} Compatibility with the biaxially fixed boundary gives a mean-stress change $\delta\sigma=-\gamma_m/V$ on switching, hence $\Delta=\gamma_m/(2\sigma_0V)$ and a noise power larger than that of a string with the same population by $(1-\nu)^{-2}$.

\emph{Loss.} The TLS couples to the oscillating bending strain of the plate through $-\gamma_m\,\mathrm{tr}\,\varepsilon=\gamma_m z\nabla^2w$, so the energy dissipated per cycle carries $\gamma_m^2$. For the volume population,
\begin{equation}
\sum_i\Delta W_i=n_v\pi\frac{h^3}{12}\frac{\gamma_m^2\langle p(1-p)\rangle}{\kB T}\Lambda\int(\nabla^2w)^2\dd A .
\end{equation}
For a clamped plate $\int(w_{xx}w_{yy}-w_{xy}^2)\dd A=0$, so the bending energy is $W_\bend=\tfrac12D\int(\nabla^2w)^2\dd A$ with $D=Eh^3/[12(1-\nu^2)]$. Dividing by $2\pi W_\bend$,
\begin{equation}
Q_{\vol,\mathrm{mem}}^{-1}=\frac{n_v\gamma_m^2\langle p(1-p)\rangle\Lambda}{E\kB T}\,(1-\nu^2)
=\frac{1+\nu}{1-\nu}\,Q_{\vol,\mathrm{string}}^{-1},
\end{equation}
larger than the string value by $(1-\nu)^{-2}(1-\nu^2)=(1+\nu)/(1-\nu)$.

\emph{Product.} Noise times $Q_\vol$ changes by $(1-\nu)^{-2}\big/[(1+\nu)/(1-\nu)]=1/(1-\nu^2)$, i.e. by $1.06$ for $\nu=0.23$. A deviatoric component of the eigenstrain couples to the shear part of the bending strain, which adds loss, but not to the mean stress, which carries the noise; it therefore lowers the ratio below $1/(1-\nu^2)$. The factor thus lies between roughly $0.65$ and $1.06$ depending on the symmetry of the eigenstrain, and we set it to unity for strings and membranes alike; the residual uncertainty of the membrane prefactor is of the same size as the scatter of the data. Strings patterned from a biaxial film have $\sigma_{0,\mathrm{string}}=(1-\nu)\sigma_{0,\mathrm{film}}$; with $\sigma_0$ extracted from the eigenfrequency of each device, as in this work, this relaxation is already contained in $\sigma_0$ and no further correction applies. A population confined to a subsurface layer of depth $d<h/2$ interpolates between the two placements of Section~\ref{sec:micro-model}, with loss weight between $1/3$ and $1$ and a count $\propto A$, and behaves as a surface population for $h\gg d$.

\section{Bending-dominated regime}\label{sec:beams}
For a beam with both ends axially fixed the compatibility argument of Eq.~\eqref{eq:dsigma} holds at any prestress. Keeping both energies,
\begin{equation}
\omega^2=\frac{W_\tens+W_\bend}{W_\kin},\quad\frac{W_\tens}{W_\bend}=\frac{\sigma_0}{\sigma_c},\quad\sigma_c=\frac{Eh^2\kappa_n^2}{12},
\end{equation}
with $\kappa_n^2=\int w''^2\dd x/\int w'^2\dd x$, equal to $(n\pi/L)^2$ for sinusoidal modes and $40.7/L^2$ for the exact clamped-clamped fundamental. Neglecting the change of mode shape with tension, $\delta(\omega^2)/\omega^2=\delta\sigma/(\sigma_0+\sigma_c)$, so
\begin{equation}
\Delta=\frac{\gamma}{2(\sigma_0+\sigma_c)V},\qquad
\hmo VQ_\vol=\frac{E\kB T}{2\pi(\sigma_0+\sigma_c)^2},
\label{eq:general}
\end{equation}
and likewise for the surface placement with $6\pi$. The loss is unaffected, and in the bending regime $Q_\mathrm{int}$ is measured directly. In the unstressed limit $\hmo V\propto\sigma_c^{-2}\propto(L/h)^4$ at fixed population.

A cantilever has no axial constraint: an eigenstrain relaxes into a static curvature and shifts the frequency only at second order. In the language of the FDT, a cantilever has no constrained tension coordinate, so the first-order thermal channel of Eq.~\eqref{main-eq:fdt} is absent and only the modulus channel of Section~\ref{sec:modulus} remains. A modulus change $\delta E_f$ at depth $z_0$ enters the bending energy with the weight of the local strain squared, $\varpi(z)=12z^2/h^2$ ($\varpi=3$ at the surface, $\langle\varpi\rangle=1$, $\langle\varpi^2\rangle=9/5$), so with $\gamma_E\equiv V_f\delta E_f$
\begin{equation}
\Delta_E=\frac{c_E\gamma_E\varpi(z_0)}{2EV},\quad
\hmo=\frac{9c_E^2K_{E,v}}{20E^2V\ln R}\ \text{(vol.)},
\end{equation}
with $K_{E,v}=n_v\langle\gamma_E^2p(1-p)\rangle$ and $c_E$ the mode-weighting factor for $w''^2$, which does not reduce to unity here because a stiffness change is local and is not redistributed by force balance. This channel has no stress dependence and, in the linear regime, no relaxation loss; the loss remains set by $\gamma$, and the product $\hmo VQ_\vol=(9c_E^2\kB T/10\pi E)\langle\gamma_E^2p(1-p)\rangle/\langle\gamma^2p(1-p)\rangle$ contains a microscopic ratio that does not cancel. In a doubly clamped beam both channels add, with $\Delta_\tens/\Delta_E=(\gamma/c_E\varpi\gamma_E)\,E/(\sigma_0+\sigma_c)$ and $E/\sigma_c\approx0.3(L/h)^2\sim10^3$ to $10^4$, so the eigenstrain channel dominates unless $\gamma_E\gg\gamma$. A doubly clamped beam and a cantilever of the same film therefore measure $\gamma_E/\gamma$.

\bibliography{refs}